\documentclass[twocolumn,showpacs,preprintnumbers,amsmath,amssymb]{revtex4}

\usepackage{graphicx}
\usepackage{dcolumn}
\usepackage{bm}

\begin{document}

\input epsf.sty



\title{Non-thermal statistics in isolated quantum spin clusters after a series of perturbations}

\author{Kai Ji}
\email{K.Ji@thphys.uni-heidelberg.de}
\affiliation{Institute for Theoretical
Physics, University of Heidelberg, Philosophenweg 19, 69120
Heidelberg, Germany}
\author{Boris V. Fine}
\email{B.Fine@thphys.uni-heidelberg.de}
\affiliation{Institute for Theoretical
Physics, University of Heidelberg, Philosophenweg 19, 69120
Heidelberg, Germany}

\date{20 June, 2011}

\begin{abstract}
We show numerically that a finite isolated cluster of interacting spins 1/2 exhibits a surprising non-thermal statistics when subjected to a series of small non-adiabatic perturbations by external magnetic field.  The resulting occupations of energy eigenstates are significantly higher than the thermal ones on both the low and the high ends of the energy spectra. This behavior semi-quantitatively agrees with the statistics predicted for the so-called ``quantum micro-canonical" (QMC) ensemble, which includes all possible quantum superpositions with a given energy expectation value. Our findings also indicate that the eigenstates of the perturbation operators are generically localized in the energy basis of the unperturbed Hamiltonian. This kind of localization possibly protects the thermal behavior in the macroscopic limit.  
\end{abstract}


\maketitle


The subject of equilibration in completely isolated quantum systems has recently seen a significant new wave of interest\cite{Casalilla-10,Kollath-07,Rigol-08,Barthel-08,Reimann-08,Zhang-10,Canovi-10,Genway-10,Pal-10,Banuls-11,Gogolin-11,Ponomarev-11,Polkovnikov-11,Ikeda-10}, which is motivated by the realization that the present-day experimental capabilities are steadily approaching the limit of probing the equilibration dynamics of isolated many-body quantum systems\cite{Kinoshita-06}. In the near future, however, the systems available for {\it decoherence-free} equilibration experiments are likely to be limited to those that have a large number of quantum levels but not too large number of particles. An isolated system of ten $q$-bits already fits into the above category. Other relevant examples are small clusters of nuclear spins, or the systems of a few cold atoms. The present paper reveals a surprising statistical behavior of such systems under the action of a series of small non-adiabatic perturbations.

If a gas of classical particles in an isolated container is subjected to a series of non-adiabatic perturbations, it is expected to return to thermal equilibrium once the perturbations stop. The condition for this equilibration is that the dynamics of the particles is sufficiently chaotic.
The quantum counterpart of the above process entails one essential complication. Namely, a non-adiabatic change of the Hamiltonian of an isolated quantum system generically broadens the participation of the energy eigenstates in the wave function of the system. Repeated non-adiabatic manipulations can therefore induce a very broad participation  of the energy eigenstates, which, in turn, may lead to a superposition of microcanonical ensembles corresponding to different temperatures\cite{Aarts-00}.
If after such manipulations the system is left to itself, the occupations of its energy eigenstates and hence the relative weights of different microcanonical ensembles will not change with time. The small subsystem behavior within such a system would be rather unconventional. 

As the above discussion suggests, the appropriate choice of the statistical ensemble depends on the preparation of a quantum system. For example, energy measurement, even if inaccurate, is likely to lead to a usual microcanonical-like ensemble\cite{Reimann-08}. However, the ensembles that might emerge in the course of unitary quantum evolution associated with non-adiabatic perturbations have not yet been fully characterized.  
This line of questioning has previously motivated one of us\cite{Fine-09-statistics,Fine-10} to investigate the so-called ``quantum micro-canonical" (QMC) ensemble\cite{Brody-98}, which is, in a sense, opposite to the conventional microcanonical ensemble. The QMC ensemble includes {\it all} quantum superpositions in an isolated quantum system that have a given energy expectation value $E_{\hbox{\scriptsize av}}$, i.e. the participating eigenstates are not limited to the narrow energy window around $E_{\hbox{\scriptsize av}}$, in contrast with the conventional microcanonical ensemble.  The QMC ensemble is also distinctly different from the conventional canonical ensemble. It leads to unconventional statistics\cite{Fine-09-statistics,Fine-10,Wootters-90,Bender-05,Jona-Lasinio-06,Brody-07A,Fresch-09,Fresch-10B,Mueller-11}, which is summarized in \cite{epaps}.

In this paper, we show that a finite isolated quantum  system with the initial state selected from the canonical ensemble permanently departs from the exponential canonical statistics after a series of small non-adiabatic perturbations. The emerging statistics semi-quantitatively agrees with the QMC statistics. Our findings also indicate that the eigenstates of the perturbation operators are generically localized in the energy basis of the unperturbed Hamiltonian. This peculiar kind of localization possibly contributes to protecting the conventional thermal behavior in the macroscopic limit. Finally, we propose that one can obtain the experimental evidence of the non-thermal statistics by performing adiabatic magnetization of spin clusters. 

\begin{figure} \setlength{\unitlength}{0.1cm}

\begin{picture}(100, 24)
{ 
\put(10, 0){ \epsfxsize= 2.5in \epsfbox{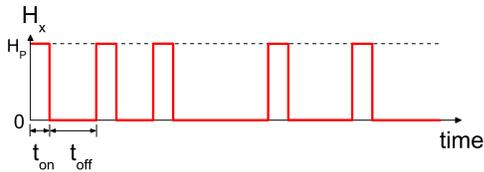} }
}
\end{picture} 
\caption{Time dependence of the external magnetic field $H_x(t)$.
} 
\label{fig-pulses} 
\end{figure}

We have performed numerical investigations of the periodic chain of 16 interacting spins 1/2 in a time-dependent external magnetic field. The Hamiltonian of the problem is
${\cal H} = {\cal H}_0 + {\cal H}_p(t)$, where 
\begin{equation}
{\cal H}_0 = \sum_{i=1}^{16} J_x S_i^x S_{i+1}^x + J_y S_i^y S_{i+1}^y + 
J_z S_i^z S_{i+1}^z ,
\label{H0}
\end{equation}
and 
${\cal H}_p(t) = H_x(t) \sum_{i=1}^{16} S_i^x$.
Here $S_i^x$, $S_i^y$, $S_i^z$  are the spin operators for the $i$th lattice site, the nearest neighbor coupling constants are $J_z = 2$ and $J_x = J_y = -1$, and $H_x(t)$ is a strong time-dependent external magnetic field along the $x$-direction, which is turned on and off in a series of rectangular pulses shown in Fig.\ref{fig-pulses}. The above choice of the coupling constants is relevant to nuclear magnetic resonance  experiments in solids (e.g. \cite{Li-08}). 


In general, the energy expectation values of quantum superpositions increase under non-adiabatic perturbations. If the perturbations are too strong, then a finite quantum system reaches the uniform infinite temperature distribution after very few such perturbations with the effect of each perturbation being largely non-universal. In a search for universal behavior, we have investigated the effect of a series of {\it small} non-adiabatic perturbations.

We introduce the non-adiabatic perturbations through the time dependence of the magnetic field $H_x(t)$ shown in Fig.\ref{fig-pulses}:  The strong magnetic field $H_P=10$ acts on the system during time interval $t_{\hbox{\scriptsize on}}$, and then it is switched off during time $t_{\hbox{\scriptsize off},n} \gg \tau$, where $\tau \sim 1$ is the characteristic timescale of one-spin motion for the Hamiltonian ${\cal H}_0$, and $n$ is the pulse number. We used the same $t_{\hbox{\scriptsize on}}$ for each pulse but have chosen $t_{\hbox{\scriptsize off},n}$ randomly between 100 and 110. As discussed later, making $t_{\hbox{\scriptsize off},n}$ constant, would make the perturbation sequence periodic, thereby leading to the dynamical localization of the occupations of the eigenstates under the perturbations.

In order to illustrate the universal trend towards the QMC statistics, we present the results for two qualitatively different kinds of pulse sequences characterized by different values of $t_{\hbox{\scriptsize on}}$. The ``short-pulse sequence" has very short $t_{\hbox{\scriptsize on}}=0.02$ such that $H_P \; t_{\hbox{\scriptsize on}} \ll 1$, which means that during time $t_{\hbox{\scriptsize on}}$ each spin rotates in magnetic field $H_P$ by a small angle. In this case, the spread of the occupations of eigenstates is expected to be of the first order in the above angle, while the drift of the average energy associated with the Hamiltonian ${\cal H}_0$ is of the second order. Therefore, initially, the energy spread is expected to grow faster than the energy drift.
The second sequence contains ``$m \pi$-pulses" characterized by $H_P \; t_{\hbox{\scriptsize on}} = m \pi$, where $m$ is an integer number. In the limit $H_P \rightarrow \infty$, an $m \pi$-pulse simply rotates each spin by $m \pi$, and hence the energy of the Hamiltonian ${\cal H}_0$ remains exactly the same. However, since, in reality, $H_P$ is finite, the $m \pi$-pulse introduces a small perturbation to the system associated with the non-commutation between ${\cal H}_0$ and ${\cal H}_p$. In comparison, with the short-pulse perturbation, the $m \pi$-pulse perturbation is expected to couple more different quantum states and hence be more effective in forcing the system to explore its Hilbert space.
The larger $m$, the stronger the perturbation due to the $m \pi$-pulse. Below, we use the sequence of $4 \pi$-pulses, for which $t_{\hbox{\scriptsize on}} = 0.4 \pi$.

We denote the wave function of the system at the end of the $n$th $t_{\hbox{\scriptsize on}}$-$t_{\hbox{\scriptsize off}}$ cycle as
\begin{equation}
\Psi_n = \sum_k C_k^{(n)} \Phi_k ,
\label{Psin}
\end{equation}
where $\Phi_k$ are the eigenfunctions of the system and $C_k^{(n)}$ the complex amplitudes.
The corresponding energy expectations values $E_{\hbox{\scriptsize av}}^{(n)}$ are
\begin{equation}
E_{\hbox{\scriptsize av}}^{(n)} = \sum_k E_k p_k^{(n)}, 
\label{Eav}
\end{equation}
where $E_k$ are the eigenenergies, and  $p_k^{(n)} \equiv |C_k^{(n)}|^2$ are occupation numbers of the respective eigenstates.

We assume that the spin cluster was initially in contact with a thermal environment characterized by temperature $T_0=1$ ($k_B = 1$) and then was slowly isolated. 
Therefore, for the initial wave function $\Psi_0$, we choose the complex amplitudes
\mbox{$C_k^{(0)} = {1 \over \sqrt{Z}} e^{{-E_k \over 2 T_0}} e^{i \alpha_k}$},
where  $\alpha_k$ are random phases, and $Z = \sum_k \hbox{exp}(- E_k/T_0)$. 
In principle, since the initial thermal state is a mixed one, the perturbations-induced ensemble should be obtained by averaging over many pure state evolutions with different random phases $\alpha_k$. However, given the large number of eigenstates, the evolution of a single pure state already contains the required statistics\cite{Linden-09,epaps}. 
The individual occupations $p_k^{(n)}$ for a single pure state strongly fluctuate between adjacent energy levels. In order to obtain the statistics of 
$p_k^{(n)}$, we organize the eigenstates into energy bins, each containing $N_L$ adjacent energy levels. For each bin, we plot $p(E)$, where $p = \sum_k^{\hbox{\scriptsize bin}} p_k^{(n)} $  and $E = {\sum_k^{\hbox{\scriptsize bin}} E_k p_k^{(n)} \over \sum_k^{\hbox{\scriptsize bin}} p_k^{(n)}} $. 

For our pulse sequences, the effect of the $n$th perturbation cycle can be expressed as 
$\Psi_n = \hat{U}_n \Psi_{n-1}$, where 
\mbox{$\hat{U}_n = \hbox{exp}\left[-i ({\cal H}_0 - H_P \sum_k S_k^x)t_{\hbox{\scriptsize on}}\right] \hbox{exp}(-i {\cal H}_0 t_{\hbox{\scriptsize off},n})$}. Starting with the ``thermal" initial wave function $\Psi_0$, we calculate the the wave functions $\Psi_n$ with the help of the complete diagonalization of the Hamiltonians ${\cal H}_0$ and ${\cal H}_0 - H_P \sum_k S_k^x$.

\begin{figure} \setlength{\unitlength}{0.1cm}

\begin{picture}(100, 56)
{ 
\put(0, 0){ \epsfxsize= 3.2in \epsfbox{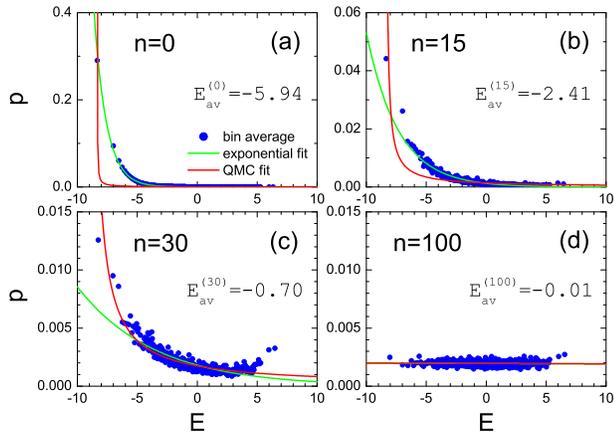} }
}
\end{picture} 
\caption{(Color online) Occupations of the eigenstates of ${\cal H}_0$ after the sequence of $n$ short pulses. Each point represents an energy bin of 128 adjacent eigenstates. The QMC and the exponential fits are calculated as described in the text. See \cite{epaps} for the semilog version of these plots.
} 
\label{fig-short} 
\end{figure}

Our numerical results for the statistics of $p_k^{(n)}$ generated by the short-pulse sequence are shown in Fig.~\ref{fig-short}. Figure~\ref{fig-short}(a) shows the initial exponential distribution. Figure~\ref{fig-short}(b) indicates how the deviations from the exponential statistics begin to develop. Figure~\ref{fig-short}(c) contains a clear case, when the distribution is much better describable by the QMC statistics than by the exponential statistics. Figure~\ref{fig-short}(d) represents the approach to the infinite temperature state, where all eigenstates are nearly equally occupied.


\begin{figure} \setlength{\unitlength}{0.1cm}

\begin{picture}(100, 55)
{ 
\put(0, 0){ \epsfxsize= 3.2in \epsfbox{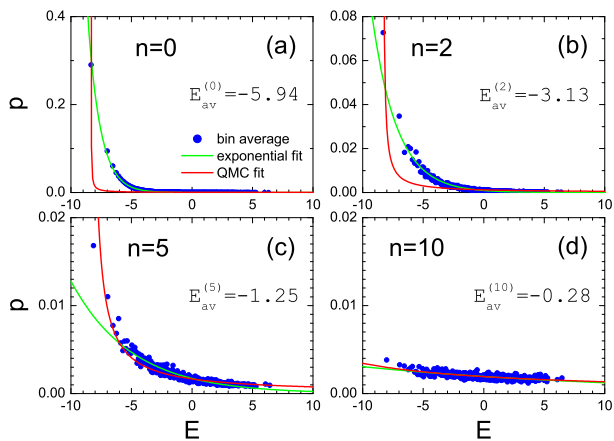} }
}
\end{picture} 
\caption{(Color online) Same as Fig.~\ref{fig-short} but for the sequence of $n$ 
$4\pi$-pulses. 
} 
\label{fig-pi} 
\end{figure}

Figure~\ref{fig-pi} presents the result of the application of the $4 \pi$-pulse sequence. In comparison with the short-pulse sequence, the $4 \pi$-pulse sequence generates the QMC distribution after much fewer pulses and with more pronounced deviations from the exponential distribution. 


We have also investigated smaller one-dimensional and two-dimensional disordered clusters, where the results were qualitatively the same\cite{epaps}.

In both Fig.~\ref{fig-short} and ~\ref{fig-pi}, we compare the numerically generated statistics with the the exponential statistics
\mbox{$p^{(n)}(E) \cong \hbox{exp} (- E/T_n)$} expected for the canonical ensemble and with the QMC statistics \mbox{$p^{(n)}(E) \cong \left[ 1+\lambda_n (E - E_{\hbox{\scriptsize av}}^{(n)}) \right]^{-1}$},
where both $T_n$ and $\lambda_n$ are determined by matching the average energies of the respective theoretical ensembles to $E_{\hbox{\scriptsize av}}^{(n)}$ --- see~\cite{epaps}.

Overall, the numerical results presented in Figs.~\ref{fig-short} and \ref{fig-pi} clearly support the notion that the ensembles generated by a series of small non-adiabatic perturbations of isolated finite spin systems approach the infinite temperature distribution through a stage, which is much better describable by the QMC statistics than by the canonical exponential statistics.
The fact that the $4 \pi$-pulse sequence generates less deviations from the QMC statistics presumably implies that its effect is closer to the action of a random rotation in the Hilbert space.

Central to understanding the emergence of the QMC statistics and the deviations from it is the competition between the spread in the occupations of the energy eigenstates and the drift of the average energy $E_{\hbox{\scriptsize av}}^{(n)}$ towards the infinite temperature value. The former reflects  the degree, to which the quantum superpositions explore the ``energy shell" in the Hilbert space associated with a given value of $E_{\hbox{\scriptsize av}}^{(n)}$ in Eq.(\ref{Eav}). The QMC ensemble represents the limit of random sampling on the above energy shell\cite{Fine-09-statistics}.  All possible eigenstates, including those on the high energy end of the spectrum, should participate in the QMC sampling, but, in order for this to happen, the effect of the perturbations should spread sufficiently fast towards the high energy eigenstates.

\begin{figure} \setlength{\unitlength}{0.1cm}

\begin{picture}(100, 45)
{ 
\put(10, -2){ \epsfxsize= 2.5in \epsfbox{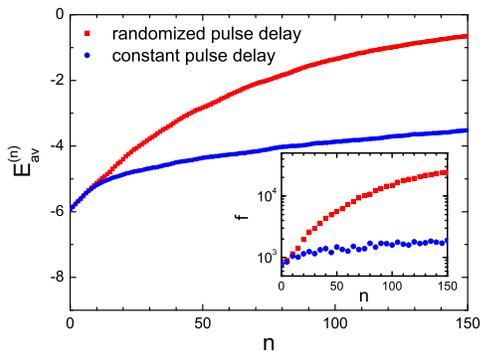} }
}
\end{picture} 
\caption{(Color online) Drift of the average energy $E_{\hbox{\scriptsize av}}^{(n)}$ for the periodic and randomized sequences of short pulses. Inset: the participation ratio of the eigenstates (see the text).
} 
\label{fig-drift} 
\end{figure}

In a related development presented in Fig.~\ref{fig-drift}, we have found that $E_{\hbox{\scriptsize av}}^{(n)}$ drifts very differently for the periodic perturbations with the constant value of the delay time $t_{\hbox{\scriptsize off},n}=100$ and for the aperiodic perturbations with the random choice of $t_{\hbox{\scriptsize off},n}$ between 100 and 110. Given that the system is expected to relax completely between two pulses separated by $t_{\hbox{\scriptsize off},n} \gg \tau$, both the periodic and the aperiodic pulse sequences could have been expected to produce the same energy drift. We came to the conclusion that the observed difference is related to the localized nature of a typical perturbation operator $\hat{U}_n$ in the energy eigenbasis of the Hamiltonian ${\cal H}_0$. If the perturbations are periodic, i.e. $\hat{U}_n = \hat{U}_1$, and if the eigenfunctions of $\hat{U}_1$ are localized as suggested, then, for $\Psi_0$ equal to one of the eigenstates of ${\cal H}_0$, no matter how many perturbations $\hat{U}_1$ one applies, the occupations $p_k$ would become non-zero only within the ``energy localization length" for the eigenstates of $\hat{U}_1$.
We believe that the incomplete saturation of the energy drift in the periodic case seen in Fig.~\ref{fig-drift} is related to the finite width of the energy spectrum, because,  in the localization regime, the transport through finite disordered systems is exponentially suppressed but still possible.
For the randomized choice of $t_{\hbox{\scriptsize off},n}$, each operator $\hat{U}_n$ is expected to possess a set of localized eigenstates, but these eigenstates should be different for different $\hat{U}_n$. Therefore, the effect of a sequence of such perturbations involves successive projections between the above localized eigenstates which leads to the continuous spread of the occupations of the eigenstates along the energy axis. The inset of Fig.~\ref{fig-drift} supports the energy localization picture by presenting the eigenstate participation ratio $f = \left[ \sum_k |C_k^{(n)}|^4 \right]^{-1}$. In the case of periodic pulses, $f$ nearly stops growing at the same time as the energy drift slows down.


The above localization property is naturally expected under the assumption that the effect of each perturbation $\hat{U}_n$ has a character of disordered one-dimensional hopping along the energy axis. The disorder in this case should originate from the complex structure of the matrix elements of $\hat{U}_n$ associated with the non-integrability of the perturbation problem. A similar kind of dynamical localization\cite{Casati-79,Fishman-82} has been investigated, e.g., in the context of the ionization of Rydberg atoms\cite{Casati-87,Buchleitner-93}.

As the number of spins $N_s$ in the system increases, we expect the drift of $E_{\hbox{\scriptsize av}}^{(n)}$ to scale linearly with $N_s$ , while the spread of the eigenstate occupation numbers is more likely to have a diffusive character  and scale as $\sqrt{N_s}$. Thus it should be more difficult to abandon the canonical ensemble for larger systems.
In the large-$N_s$ limit, the eigenstate occupation profiles associated with the above-mentioned energy localization may further contribute to protecting the conventional thermal behavior\cite{epaps}.

If a QMC-like statistics is generated experimentally in a cluster of $N_s$ spins 1/2, how can one detect it? A possibly effective way of doing it is to
perform the adiabatic magnetization of the system with the subsequent measurement of the total magnetization. The above procedure is implemented numerically in \cite{epaps}.

In conclusion, we have presented numerical evidence for emerging QMC-like statistics in isolated spin clusters after non-adiabatic perturbations and for the dynamical localization property of the perturbation operators. Both of the above properties need to be further investigated in terms of their dependence on the type and the size of quantum systems, on the perturbation routines and also in terms of their stability in the presence of decoherence and subsequent quantum measurements.

We are grateful to the bwGRiD computing cluster and to the Innovation Fund Frontier at the University of Heidelberg for supporting this work. 

{\it Note added:} A very recent preprint of Bunin et al.\cite{Bunin-11} considers the issue of energy drift vs. eigenstate occupations spread in a similar context.


\begin{center}
{\bf SUPPLEMENTARY MATERIAL }
\end{center}

\setcounter{figure}{0}
\renewcommand{\thefigure}{S\arabic{figure}}

\setcounter{equation}{0}
\renewcommand{\theequation}{S\arabic{equation}}


Note: Reference numbers in the text below are from the reference list of the main article.

\subsection{Quantum micro-canonical ensemble.}
\label{QMC}

Quantum micro-canonical (QMC) ensemble is defined for $N$-dimensional Hilbert spaces to include all possible superpositions of eigenstates 
\begin{equation}
\Psi = \sum_{k=1}^{N} C_k \Phi_k
\label{Psi}
\end{equation}
with a given energy expectation value $E_{\hbox{\scriptsize av}}$, i.e.
\begin{equation}
\sum_{i=k}^{N}  E_k |C_k|^2  = E_{\hbox{\scriptsize av}}.
\label{epsav}
\end{equation}
``All possible superpositions'' means uniform joint probability measure on the $(2N-2)$-dimensional manifold constrained by condition (\ref{epsav}) and by the normalization condition $\sum_{i=k}^{N}  |C_k|^2  =1$ in the $2N$-dimensional Euclidean space of variables $\{ \hbox{Re}C_k, \hbox{Im}C_k \}$. 

The QMC ensemble allows no exchange of energy or particles with the environment. Therefore, at first sight, it appears to be similar to the conventional micorcanonical ensemble. The crucial difference, however, is that the conventional microcanonical ensemble, when applied to quantum systems, limits the participation of the energy eigenstatates to a narrow energy window of unspecified small width (which however, should be large enough to include many eigenstates). In contrast, the QMC ensemble allows all eigenstates to participate as long as the low-energy eigenstates balance the high-energy ones to satisfy condition (\ref{epsav}). The broad participation of energy eigenstates in the QMC ensemble may also appear to be similar to the canonical ensemble. Here, however, clear quantitative differences exist, which, for example, can be seen in Figs.~2 and 3 of the main article.

The statistical properties of the QMC ensemble is more convenient to discuss in terms of
the occupation-of-eigenstate variables $p_k = |C_k|^2$ and the phases of quantum amplitudes $\alpha_k$ defined by $C_k = |C_k| e^{i \alpha_k}$. The two QMC  constraints can then be expressed as 
\begin{equation}
\sum_{i=k}^{N}  E_k p_k  = E_{\hbox{\scriptsize av}},
\label{epsav1}
\end{equation}
and
\begin{equation}
\sum_{i=k}^{N}  p_k  = 1.
\label{norm}
\end{equation}
These constraints impose no restriction on phases $\alpha_k$, which, therefore, can be chosen completely randomly in the interval $[0, 2\pi)$. It can also be shown that the uniform joint probability measure on the QMC manifold in the space of variables $\{ \hbox{Re}C_k, \hbox{Im}C_k \}$ translates into the uniform joint probability measure in the space of variables $\{ p_k \}$.

Application of the QMC ensemble requires two steps: (I) Calculation of the marginal probability distribution  for the occupation numbers $p_k$, which we denote as $P_k (p_k)$, and which gives the corresponding average value $\langle p_k \rangle$; and (II) Calculation of the density matrix for a small subsystem. Both steps where done analytically for large $N$ in Ref.[17], and the finite-$N$ corrections were introduced in Ref.[18].  The numerical results presented in the present work focus on step (I) and indicate that the statistics for $\langle p_k \rangle$ generated by a series of non-adiabatic perturbations becomes close to QMC.  (In the case of the conventional canonical and microcanonical ensembles,  the values of $\langle p_k \rangle$ are simply postulated.) Step II is discussed at the end of this section.

The QMC ensemble in the small-$p_k$ approximation (the most general case) leads to the following marginal probability distribution[17]
\begin{equation}
P_k(p_k) \cong  e^{-N p_k[1 + \lambda(E_k - E_{\hbox{\scriptsize av}})]}.
\label{Pk}
\end{equation}
As a result,
\begin{equation}
\langle p_k \rangle = {1 \over N [1 + \lambda (E_k - E_{\hbox{\scriptsize av}})]}.
\label{pav}
\end{equation}
Here $\lambda$ is a parameter that can be determined by substituting Eq.(\ref{pav}) into the averaged versions of either Eq.(\ref{epsav1}) or Eq.(\ref{norm}) and then solving the resulting equation for $\lambda$ numerically.
It should be noted that the exponentially decaying distribution $P_k(p_k)$ has maximum not at $p_k=\langle p_k \rangle$ but rather at $p_k=0$, which means that the individual values of $p_k$ for any particular superposition $\Psi$ belonging to the QMC ensemble strongly fluctuate with respect to $\langle p_k \rangle$. For this reason, in analyzing the numerical results, we use the energy binning procedure, which groups together $N_L$ adjacent levels having very close values of energies $E_k$ and, therefore, according to Eq.(\ref{Pk}), nearly the same probability distributions $P_k(p_k)$. We convert formula (\ref{pav}) into the following approximation for the total occupation of the quantum states within the bin at energy  $E$
\begin{equation}
p(E) = {N_L \over N [1 + \lambda (E - E_{\hbox{\scriptsize av}})]},
\label{pavbin}
\end{equation}
and then calculate $\lambda$ numerically by substituting $p(E)$ into the the ``binned'' version of Eq.(\ref{epsav1}).

One may be concerned that the integrals $\int p(E) dE $ and $\int E p(E) dE$ with $p(E)$ given by Eq.(\ref{pavbin}) appear to diverge at large $E$. In this regard, it is necessary to remember that the QMC ensemble is defined for large but finite Hilbert spaces which always have the maximum value of $E$ terminating the integration range.
As the size of the system increases, the high energy tail of $p(E)$ needs to be radically balanced by the occupations of the low energy levels to deliver the given value of $E_{\hbox{\scriptsize av}}$. It was shown in Ref.[17], that, for the systems of macroscopic size, the QMC ensemble would lead to a generic condensation into the ground state of the system. (See sections~II.E and II.F of Ref.[17] for more details.)
In the present work, our numerically generated QMC-like ensemble is far from the above condensation regime.

The perturbations of the external magnetic field in our simulations drive the system towards the infinite temperature equilibrium. In this work, however, we are not interested in what happens after infinitely many perturbations. The question we ask is: what happens if, after $n$ perturbations, the system is left completely isolated. A classical system in this case would thermally equilibrate, because it would be describable by a microcanonical ensemble with the value of energy reached after the last perturbation.
In the quantum case, however, as we demonstrated in this work, the resulting ensemble is QMC-like for a range of values of $n$. Since the occupations of eigenstates of isolated quantum system do not change with time, the resulting QMC-like ensemble will not evolve further, which, in turn, implies that the stationary density matrices  of small subsystems within such an isolated system will deviate from the predictions based on the conventional thermal equilibrium. 

In this work, we use many {\it small} perturbations, which change the energy expectation value of a quantum superposition only by small amount but, at the same time, force the system to sample the Hilbert space.
We suppose that our perturbation sequences generate the QMC-like statistics, because the the QMC statistics describes the properties of the overwhelmingly typical quantum superposition with a given energy expectation value, in very much the same way as the Boltzmann-Gibbs distribution represents the statistics of an energy shell in the classical phase space. The difference between the quantum and the classical cases is that in the classical case, the phase space trajectory of an isolated system explores the energy shell dynamically, while, in the quantum case it does not do it by itself - only the perturbations force the system to do it, albeit with a simultaneous increase of energy.
After sufficiently many perturbations and for not too large quantum systems, we expect the quantum state of our system to become representative of the ``Hilbert space energy shell'' with the appropriate energy expectation value.  A representative sampling of Hilbert space energy shells by the technique of small perturbations may encounter practical limitations as the number of particles in the system increases, because, in this case, the volume change of the Hilbert space energy shell after each small perturbation should increase exponentially. As indicated in the main article, this question requires further investigation.

All characteristics of small subsystems, such as density matrix or correlation functions, involve integrals containing the product $\nu(E) p(E)$, where $\nu(E)$ is the density of states for the entire system. The Gibbs equilibrium for small subsystems requires that $\nu(E) p(E)$ is narrowly peaked around $E_{\hbox{\scriptsize av}}$[18]. This is always the case for the microcanonical ensemble and is also the case for the canonical ensemble in the macroscopic limit[18]. At the same time, the QMC ensemble in the macroscopic limit leads to a double-peak structure of $\nu(E) p(E)$: one peak corresponds to the above-mentioned condensation to the ground state, while the other peak corresponds to the infinite temperature state associated with the maximum of $\nu(E)$. (In this work, the energy corresponding to the infinite temperature is $E=0$.) For the finite systems considered in this work, the product of $\nu(E) p(E)$ is not narrowly peaked for the canonical ensemble. Neither it exhibits two narrow peaks for the QMC ensemble.
In such a case, distinguishing the QMC from the canonical ensemble at the level of a subsystem is possible but more involved because of the non-universal features associated with the finite size effects for either of these two ensembles. We plan to discuss the properties of subsystems elsewhere. 

Finally, we would like to remark, that the energy localization property reported in this manuscript may lead to an exponentially decaying tails of the occupation of eigenstates even for the randomized pulse sequences.  In the macroscopic limit and on the higher-energy end of $p(E)$, this exponentially decaying tail would look like the canonical ensemble, thereby leading to the narrowly peaked shape of $\nu(E) p(E)$. Such a shape, in turn, would justify the Gibbs equilibrium for small subsystems.

\subsection{Numerical results for small disordered spin clusters.}
\label{disordered}

The Hamiltonians ${\cal H}_0$ and ${\cal H}_p(t)$ considered in the main article are translationally invariant. In addition, in the quantization basis of operators $S_i^x$, neither ${\cal H}_0$ nor ${\cal H}_p(t)$ has off-diagonal elements coupling the quantum states having even and odd numbers of spins with positive $x$-projections. As a result, the quantum basis can be decomposed into 32 non-mixing subspaces, each associated with one of the 16 possible values of the inverse lattice wave vectors and with the even/odd property. Below we break both of the above symmetries by introducing additional disorder. In comparison with the 16-spin chain considered in the main article, the disordered systems investigated below have smaller numbers of spins but larger sizes of mixing Hilbert spaces. These examples are included also to illustrate that the Bethe-ansatz intergrability of the Hamiltonian ${\cal H}_0$ for spin chains is not a determining factor in the emergence of the QMC-like statistics.

In this section, we modify the Hamiltonian ${\cal H}_0$ as follows:
$$
{\cal H}_0 = \sum_{i<j}^{\hbox{\scriptsize NN}} J_x S_i^x S_j^x + J_y S_i^y S_j^y + 
J_z S_i^z S_j^z + \sum_i h_i^z S_i^z,
$$
In comparison with Eq.(1) of the main article, the lattice indices in the interaction part are modified to accommodate the nearest neighbor interaction (NN) for an arbitrary lattice dimensionality, but the main difference is the last term. It contains random local field $h_i^z$ along the $z$-direction. In the context of NMR experiments, this term may represent the disorder in the values of chemical shifts.  All other parameters of the simulations remain the same as in the main article unless indicated. [Note: In the presence of the above disorder, only $m\pi$-pulses with even $m$ perfectly conserve the average energy associated with ${\cal H}_0$ in the limit $H_P \rightarrow \infty$.]

\begin{figure} \setlength{\unitlength}{0.1cm}

\begin{picture}(100, 55)
{ 
\put(0, 0){ \epsfxsize= 3.2in \epsfbox{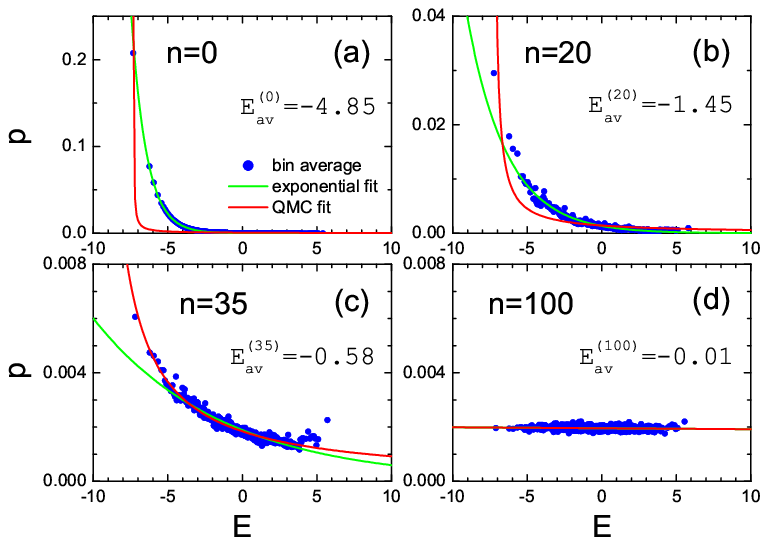} }
}
\end{picture} 
\caption{Occupations of the eigenstates of ${\cal H}_0$ for a 13-spin chain with disorder after a sequence of $n$ short pulses. Each point represents an energy bin of 16 adjacent eigenstates with additional averaging over 48 independent choices of the initial random phases $\alpha_k$. The QMC and the exponential fits are calculated as described in the main article.
} 
\label{fig-short-13} 
\end{figure}

\begin{figure} \setlength{\unitlength}{0.1cm}

\begin{picture}(100, 60)
{ 
\put(0, 0){ \epsfxsize= 3.2in \epsfbox{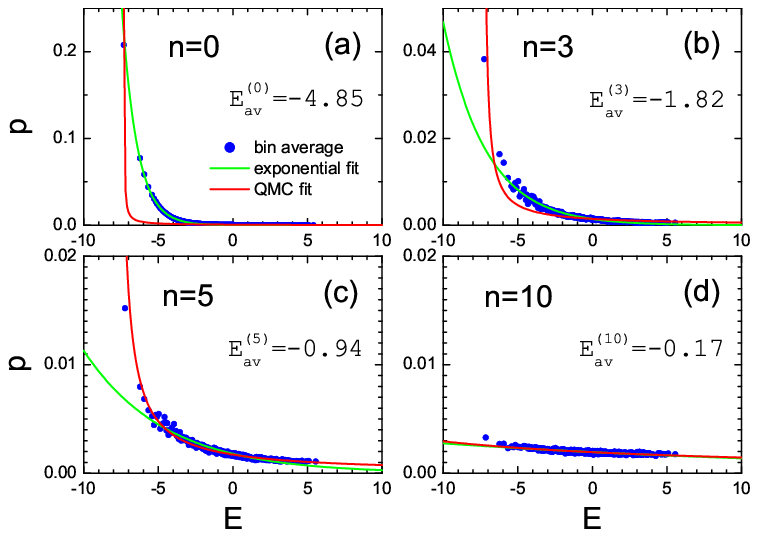} }
}
\end{picture} 
\caption{Same as Fig.~\ref{fig-short-13} but for a sequence of $n$ $4\pi$-pulses. 
} 
\label{fig-pi-13} 
\end{figure}

In Figs.~\ref{fig-short-13} and \ref{fig-pi-13}, we present the results for the short-pulse and the $4 \pi$-pulse sequences, respectively, acting on a periodic 13-spin with local fields $h_i^z$ chosen randomly in the interval $[-0.2,0.2]$. In the both figures, the results are additionally averaged over 48 independent random choices of the phases $\{ \alpha_k \}$ for the initial wave function $\Psi_0$. This additional averaging reduces the statistical noise, but as mentioned in the main article, the energy-binned result for a single randomly chosen set of $\{ \alpha_k \}$ already exhibits the same average occupations of eigenstates. 


\begin{figure} \setlength{\unitlength}{0.1cm}

\begin{picture}(100, 55)
{ 
\put(0, 0){ \epsfxsize= 3.2in \epsfbox{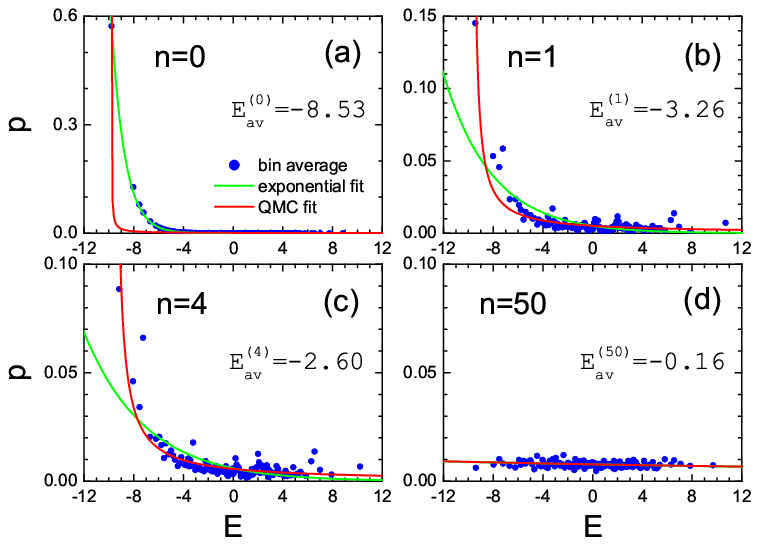} }
}
\end{picture} 
\caption{Occupations of the eigenstates of ${\cal H}_0$ for a $4 \times 3$ square lattice with disorder after a sequence of $n$ $2\pi$-pulses. Each point represents an energy bin of 32 adjacent eigenstates. The QMC and the exponential fits are calculated as described in the main article.
} 
\label{fig-pi-4x3} 
\end{figure}

In Fig.~\ref{fig-pi-4x3}, we present the results for the sequence of $2\pi$-pulses acting on a periodically closed $4 \times 3$ square lattice with local fields $h_i^z$ chosen randomly in the interval $[-0.1,0.1]$.

\subsection{Series of $4\pi$ pulses  for periodic spin chains of different length}

\begin{figure} \setlength{\unitlength}{0.1cm}

\begin{picture}(100, 60)
{ 
\put(0, 0){ \epsfxsize= 3.2in \epsfbox{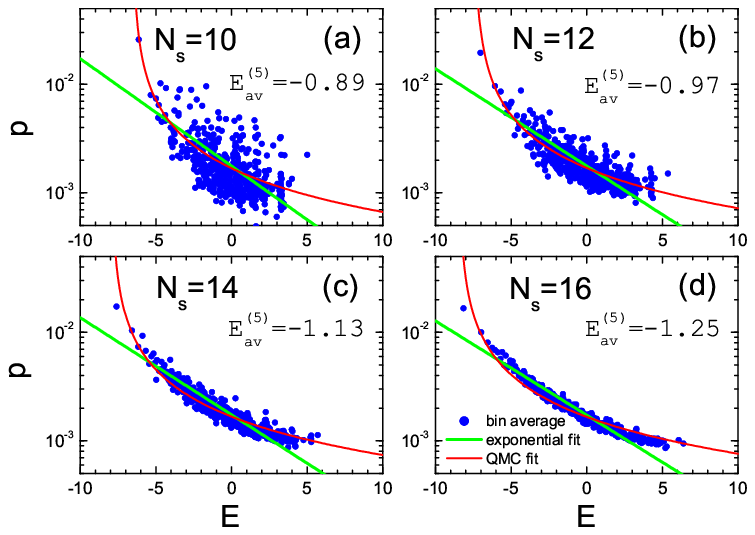} }
}
\end{picture} 
\caption{Occupations of the eigenstates of ${\cal H}_0$ for periodic spin chains of different length $N_s$ after a sequence of five $4\pi$-pulses. Each blue point represents one energy bin of adjacent eigenstates. The eigenstates in each case were divided into 512  equal energy bins.  The QMC and the exponential fits are calculated as described in the main article.
} 
\label{fig-pi-Ns} 
\end{figure}

In Fig.~3 of the main article, we present the statistics for the occupations of eigenstates generated by different numbers of $4 \pi$-pulses on the 16-spin periodic spin chain with anisotropic interaction. Here we fix the number of $4 \pi$-pulses ($n=5$) while changing the number of spins $N_s$ in the spin chain from 10 to 16. The initial ensemble in each case is chosen in the same way as in the main article --- it corresponds to the thermal occupations of the eigenstates with temperature $T=1$. The results are presented in Fig.~\ref{fig-pi-Ns}. 

\subsection{Adiabatic magnetization}

Here we describe how one can detect the QMC statistics using the  adiabatic magnetization routine mentioned in the main article.

If the magnetic field $H_x$ is increased sufficiently slowly, the eigenstates evolve, their energy change, but their occupations do not change much. At a large enough value of the magnetic field, the energy of each eigenstate is mainly due to the Zeeman energy of the total magnetization in the magnetic field. By adiabatically magnetizing a single cluster, and then doing the measurement of its total polarization $M_x = \sum_i S_i^x$, one can obtain $F_{M_x}$, the probability for the occurrence of different values of $M_x$. The number of the quantum states with a given value $M_x$ is $d_{M_x} = {N_s! \over (1/2 N_s - M_x)! \; (1/2 N_s + M_x)! }$. Therefore, the occupation of each of these states is $P_{M_x} = F_{M_x}/ d_{M_x}$. The values of $P_{M_x}$ are expected to exhibit nearly the same statistics as the occupations of the cluster eigenstates before to the magnetization begins.

\begin{figure}[h] \setlength{\unitlength}{0.1cm}

\begin{picture}(100, 58)
{ 
\put(0, 0){ \epsfxsize= 3.0in \epsfbox{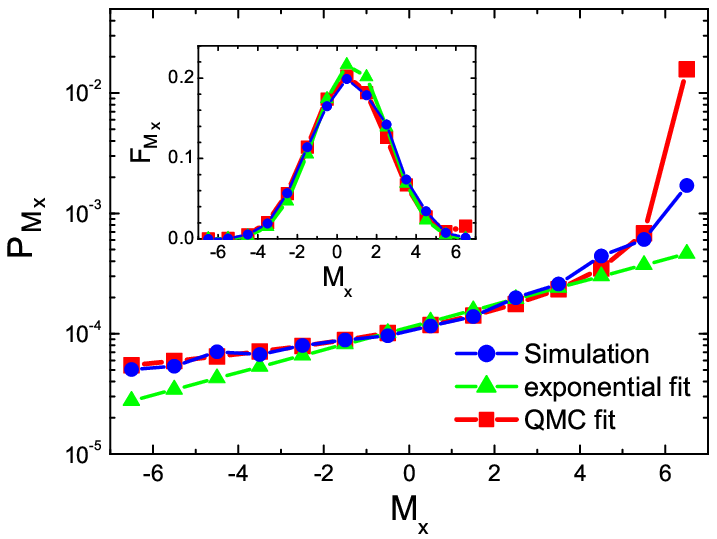} }
}
\end{picture} 
\caption{Occupations of individual quantum states (blue circles) with a given value of spin cluster polarization $M_x$ after numerical imitation of adiabatic magnetization for a 13-spin chain with small disorder. The adiabatic magnetization process started from the eigenstate occupations shown in Fig.~\ref{fig-pi-13}(c). The green triangles and the red squares represent, respectively, the exponential and the QMC fits obtained as described in the text. The inset shows  the probability of measuring a given value of $M_x$ for the simulation results (blue dots), exponential fit (green triangles) and the QMC fit (red squares).  
} 
\label{fig-adiabatic} 
\end{figure}

We have performed a numerical imitation of  the above procedure.  We started from the numerically generated QMC-like  state of the 13-spin chain shown in Fig.~\ref{fig-pi-13}(c). The magnetization was performed by increasing the magnetic field $H_x$ from 0 to 10 in steps of 0.1 separated by the waiting time 1000. The result is shown in Fig.~\ref{fig-adiabatic}. The signature of the QMC statistics in such an imperfect adiabatic procedure is rather clear.
One expects $P_{M_x} \cong \hbox{exp}(M_x H_x/T^{\prime})$, if the magnetization process starts from the canonical distribution, and
$P_{M_x} \cong [1 - \lambda^{\prime} H_x (M_x - M_{x,\hbox{\scriptsize av}})]^{-1}$, if the magnetization starts from the QMC ensemble. Here $M_{x,\hbox{\scriptsize av}} = \sum_{M_x} F_{M_x} M_x$ evaluated for the numerically generated ensemble. For the respective fits shown in Fig.~\ref{fig-adiabatic}, both $T^{\prime}$ and $\lambda^{\prime}$ were determined numerically from the simulations-generated value of $M_{x,\hbox{\scriptsize av}}$.

The inset of Fig.~\ref{fig-adiabatic} shows the probability $F_{M_x}$ of measuring a given value of $M_x$. This plot indicates that the measurement outcomes would be dominated by the states around $M_x=0$ due to the large degeneracy $d_{M_x}$ of these states. Therefore, a large number of single cluster measurements should be made in order to extract the signatures of the QMC statistics around the minimum and the maximum values of $M_x$. If many equivalent electronic spin clusters are available for simultaneous manipulations, then one may try to bring nuclear spins in contact with these clusters and then measure the distribution of $M_x$ through its effect on the NMR lineshape.


\subsection{Semi-logarithmic plots corresponding to Figs.~2 and 3 of the main article and to Figs.~\ref{fig-short-13}, \ref{fig-pi-13} and \ref{fig-pi-4x3} of the supplement.}


\begin{figure}[h]  \setlength{\unitlength}{0.1cm}

\begin{picture}(100, 55)
{ 
\put(0, 0){ \epsfxsize= 3.2in \epsfbox{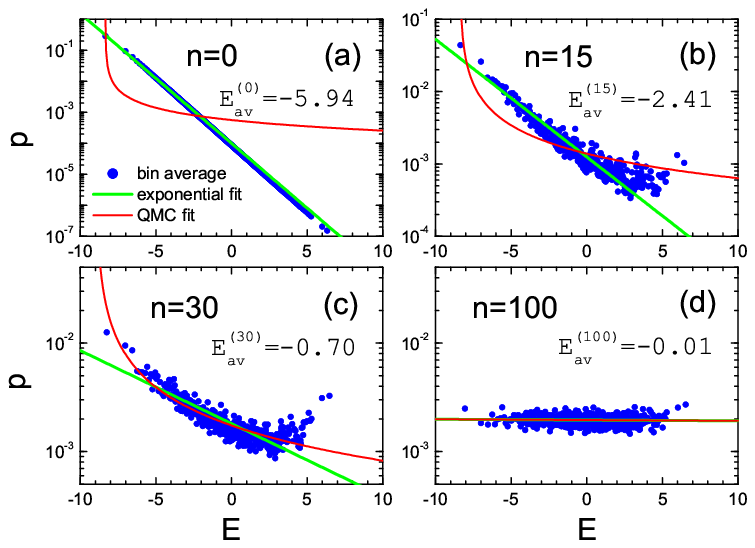} }
}
\end{picture} 
\caption{Semilog version of the plots shown in Fig.~2 of the main article. The plots represent the results of the short-pulse sequence acting on a 16-spin chain without disorder.
} 
\label{fig-short-Log} 
\end{figure}


\


\begin{figure}  \setlength{\unitlength}{0.1cm}

\begin{picture}(100, 55)
{ 
\put(0, 0){ \epsfxsize= 3.2in \epsfbox{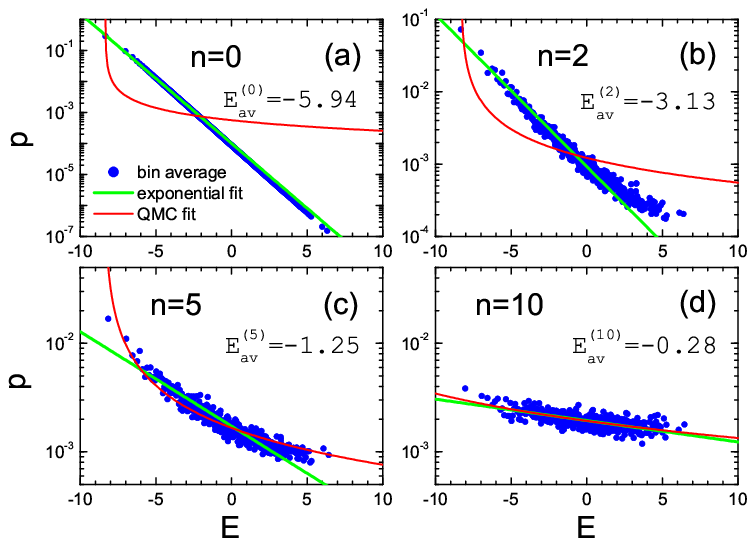} }
}
\end{picture} 
\caption{Semilog version of the plots shown in Fig.~3 of the main article. The plots represent the results of the $4 \pi$-pulse sequence acting on a 16-spin chain without disorder.
} 
\label{fig-pi-Log} 
\end{figure}


\


\begin{figure}  \setlength{\unitlength}{0.1cm}

\begin{picture}(100, 60)
{ 
\put(0, 0){ \epsfxsize= 3.2in \epsfbox{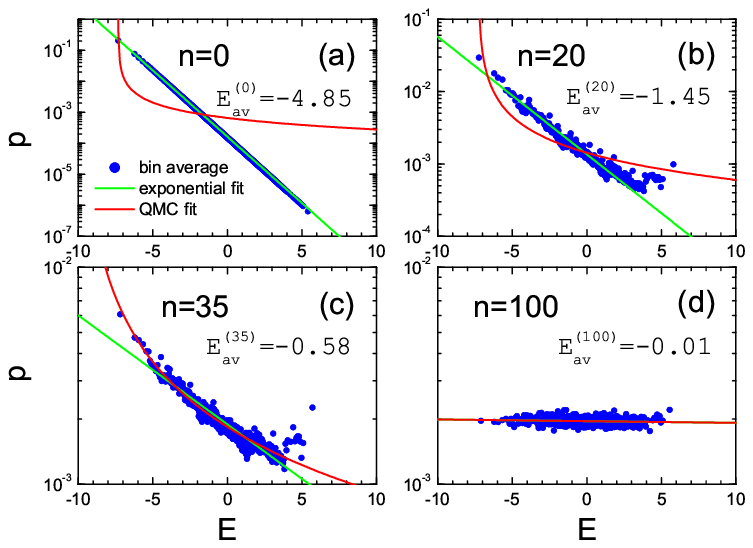} }
}
\end{picture} 
\caption{Semilog version of the plots shown in Fig.~\ref{fig-short-13}. The plots represent the results of the short-pulse sequence acting on a 13-spin chain with disorder.
} 
\label{fig-short-13-Log} 
\end{figure}


\


\begin{figure} \setlength{\unitlength}{0.1cm}

\begin{picture}(100, 55)
{ 
\put(0, 0){ \epsfxsize= 3.2in \epsfbox{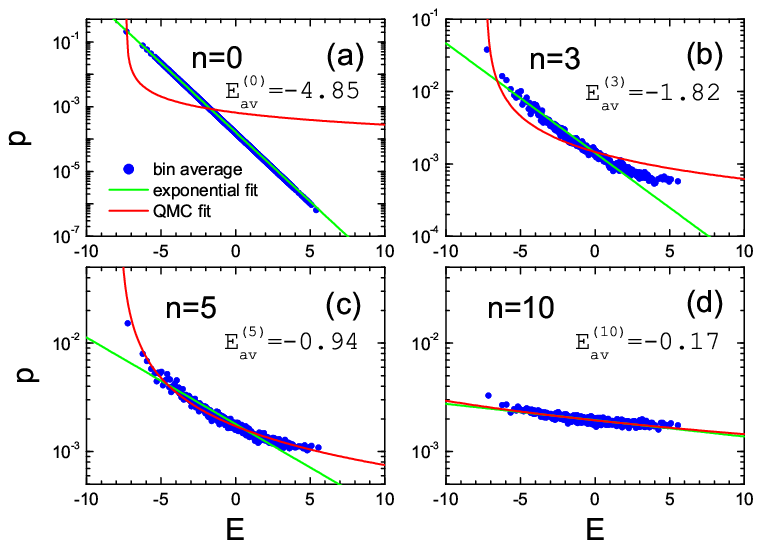} }
}
\end{picture} 
\caption{Semilog version of the plots shown in Fig.~\ref{fig-pi-13}. The plots represent the results of the $4\pi$-pulse sequence acting on a 13-spin chain with disorder.
} 
\label{fig-pi-13-Log} 
\end{figure}


\


\begin{figure} \setlength{\unitlength}{0.1cm}

\begin{picture}(100, 55)
{ 
\put(0, 0){ \epsfxsize= 3.2in \epsfbox{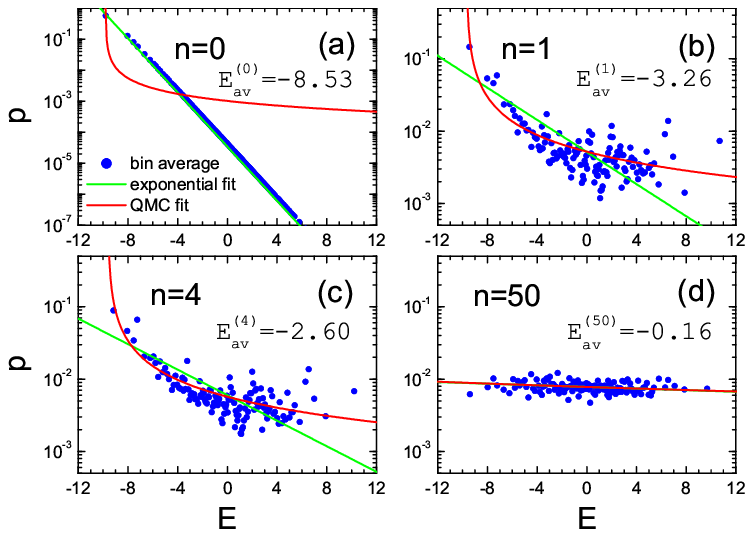} }
}
\end{picture} 
\caption{Semilog version of the plots shown in Fig.~\ref{fig-pi-4x3}. The plots represent the results of the $2\pi$-pulse sequence acting on a $4 \times 3$ square spin lattice with disorder.
} 
\label{fig-pi-4x3-Log} 
\end{figure}


\clearpage

\end{document}